\documentclass[a4paper,fleqn,usenatbib]{mnras}

\usepackage[T1]{fontenc}
\usepackage{ae,aecompl}

\usepackage{graphicx}	
\usepackage{amsmath}	
\usepackage{amssymb}	

\def\equationautorefname~#1\null{Equation~(#1)\null}
\newcommand{\msun}{\hbox{$\mathrm{M}_\odot$}}
\newcommand{\re}{\hbox{$R_{\rm e}$}}
\newcommand{\atl}{ATLAS$^{\rm 3D}$}
\newcommand{\hi}{{\sc H\,i}}
\newcommand{\kms}{km s$^{-1}$}
\newcommand{\se}{\hbox{$\sigma_\mathrm{e}$}}
\newcommand{\vhi}{\hbox{$v_\mathrm{circ}$(\hi)}}
\newcommand{\rhi}{$R_\mathrm{HI}$}
\newcommand{\rmax}{$R_\mathrm{max}$}
\newcommand{\vjam}{\hbox{$v_\mathrm{circ}^\mathrm{max}$(JAM)}}
\newcommand{\vcirc}{\hbox{$v_\mathrm{circ}$}}

\defcitealias{cappellari2013a}{C13}
\defcitealias{cappellari2015}{C15}
\defcitealias{denheijer2015}{dH15}
\defcitealias{pizzella2005}{P05}
\defcitealias{courteau2007}{C07}

\title[\vhi-\se\ relation and density profiles in ETGs]{Linear relation between \hi\ circular velocity and stellar velocity dispersion in early-type galaxies, and slope of the density profiles}

\author[Serra et al.]{\parbox{\textwidth}{
Paolo Serra,$^{1}$\thanks{E-mail: paolo.serra@csiro.au}
Tom Oosterloo,$^{2,3}$
Michele Cappellari,$^{4}$
Milan den Heijer,$^{5,6}$
and Gyula~I.~G.~J\'{o}zsa$^{5,7,8}$
}
\vspace{0.4cm}\\ 
\parbox{\textwidth}{
$^{1}$CSIRO Astronomy and Space Science, Australia Telescope National Facility, PO Box 76, Epping, NSW 1710, Australia\\
$^{2}$Netherlands Institute for Radio Astronomy (ASTRON), Postbus 2, 7990 AA Dwingeloo, The Netherlands\\
$^{3}$Kapteyn Astronomical Institute, University of Groningen, Postbus 800, 9700 AV Groningen, The Netherlands\\
$^{4}$Sub-department of Astrophysics, Department of Physics, University of Oxford, Denys Wilkinson Building, Keble Road, Oxford, OX1~3RH, UK\\
$^{5}$Argelander Institut f\"{u}r Astronomie (AIfA), University of Bonn, Auf dem H\"{u}gel 71, 53121 Bonn, Germany\\
$^{6}$Max-Planck Institut f\"{u}r Radioastronomie (MPIfR), Auf dem H\"{u}gel 69, 53121 Bonn, Germany\\
$^{7}$SKA South Africa, Radio Astronomy Research Group, 3rd Floor, The Park, Park Road, Pinelands 7405, South Africa\\
$^{8}$Rhodes University, Department of Physics and Electronics, Rhodes Centre for Radio Astronomy Techniques \& Technologies, PO Box 94, Grahamstown 6140, South Africa
}}

\date{Accepted 2016 April 26. Received 2016 April 26; in original form 2016 March 6}

\pubyear{2015}

\begin{document}
\label{firstpage}
\pagerange{\pageref{firstpage}--\pageref{lastpage}}
\maketitle

\begin{abstract}
We report a tight linear relation between the \hi\ circular velocity measured at 6 \re\ and the stellar velocity dispersion measured within 1 \re\ for a sample of 16 early-type galaxies with stellar mass between $10^{10}$ and $10^{11}$ \msun. The key difference from previous studies is that we only use spatially resolved \vhi\ measurements obtained at large radius for a sizeable sample of objects. We can therefore link a kinematical tracer of the gravitational potential in the dark-matter dominated outer regions of galaxies with one in the inner regions, where baryons control the distribution of mass. We find that \vhi~=~1.33~\se\ with an observed scatter of just 12 percent. This indicates a strong coupling between luminous and dark matter from the inner- to the outer regions of early-type galaxies, analogous to the situation in spirals and dwarf irregulars. The \vhi-\se\ relation is shallower than those based on \vcirc\ measurements obtained from stellar kinematics and modelling at smaller radius, implying that \vcirc\ declines with radius -- as in bulge-dominated spirals. Indeed, the value of \vhi\ is typically 25 percent lower than the maximum \vcirc\ derived at $\sim0.2\ R_\mathrm{e}$ from dynamical models. Under the assumption of power-law total density profiles $\rho \propto r^{-\gamma}$, our data imply an average logarithmic slope $\langle\gamma\rangle=2.18\pm0.03$ across the sample, with a scatter of 0.11 around this value. The average slope and scatter agree with recent results obtained from stellar kinematics alone for a different sample of early-type galaxies.
\end{abstract}




\section{Introduction}
\label{sec:intro}

\begin{table*}
{\centering
\caption{\hi-rich early-type galaxies}
\label{table:tab1}
\begin{tabular}{cccccccc}
\hline
Name & \re & \se & \rmax & \vjam & \rhi & \vhi & $\gamma$ \\
 & (arcsec) & (\kms)  & (arcsec) & (\kms)  & (arcsec) & (\kms) &\\
(1) & (2) & (3) & (4) & (5) & (6) & (7) & (8) \\
\hline
NGC~2685 & 22.1 & $ 104 $ & 4.0   & $ 163 $ & 320 & $ 144 \pm 10 $ &  $2.06 \pm 0.04$ \\
NGC~2824 & 8.0   & $ 127 $ & 1.9   & $ 277 $ & 40    & $ 162 \pm 10 $ &  $2.36 \pm 0.05$ \\
NGC~2859 & 27.6 & $ 163 $ & 6.6   & $ 305 $ & 115 & $ 215 \pm 41 $ &  $2.25 \pm 0.14$ \\
NGC~2974 & 27.6 & $ 226 $ & 7.9   & $ 369 $ & 130 & $ 310 \pm 10 $ &  $2.12 \pm 0.04$ \\
NGC~3522 & 14.0 & $ 98   $ & 2.5   & $ 187 $ & 85    & $ 121 \pm 8 $ &  $2.25 \pm 0.05$ \\
NGC~3626 & 24.6 & $ 131 $ & 3.3   & $ 248 $ & 120 & $ 169 \pm 8 $ &  $2.21 \pm 0.04$ \\
NGC~3838 & 9.4   & $ 133 $ & 3.5   & $ 231 $ & 150 & $ 159 \pm 14 $ &  $2.20 \pm 0.05$ \\
NGC~3941 & 24.9 & $ 121 $ & 4.5   & $ 210 $ & 195 & $ 148 \pm 8 $ &  $2.18 \pm 0.04$ \\
NGC~3945 & 29.7 & $ 177 $ & 9.1   & $ 342 $ & 130 & $ 237 \pm 13 $ &  $2.28 \pm 0.06$ \\
NGC~3998 & 24.0 & $ 224 $ & 7.2   & $ 435 $ & 195 & $ 246 \pm 20 $ &  $2.35 \pm 0.06$ \\
NGC~4203 & 38.5 & $ 129 $ & 6.2   & $ 222 $ & 195 & $ 197 \pm 35 $ &  $2.07 \pm 0.11$ \\
NGC~4262 & 11.6 & $ 161 $ & 3.8   & $ 366 $ & 120 & $ 198 \pm 10 $ &  $2.36 \pm 0.04$ \\
NGC~4278 & 33.4 & $ 213 $ & 11.4 & $ 364 $ & 150 & $ 256 \pm 26 $ &  $2.27 \pm 0.09$ \\
NGC~5582 & 28.9 & $ 148 $ & 5.0   & $ 262 $ & 210 & $ 258 \pm 10 $ &  $2.01 \pm 0.03$ \\
NGC~6798 & 12.4 & $ 130 $ & 5.0   & $ 223 $ & 150 & $ 190 \pm 8 $ &  $2.10 \pm 0.04$ \\
UGC~06176 & 9.7 & $ 96   $ & 0.9   & $ 209 $ & 60    & $ 144 \pm 14 $ &  $2.18 \pm 0.05$ \\
\hline
\end{tabular}
}

\it Column~1\rm : galaxy name; \it Column~2\rm : circularised projected half-light radius from \citetalias{cappellari2013a}; \it Column~3\rm : stellar velocity dispersion measured from integral-field spectroscopy within the half-light ellipse (uncertainty $\sim5$ percent; \citetalias{cappellari2013a}); \it Column~4\rm : radius of the peak circular velocity of the JAM models (\citetalias{cappellari2013a}); \it Column~5\rm : peak circular velocity of the JAM models (uncertainty $\sim5$ percent; \citetalias{cappellari2013a}); \it Column~6\rm : radius where \citetalias{denheijer2015} measure the \hi\ circular velocity; \it Column~7\rm : \hi\ circular velocity  (\citetalias{denheijer2015}); \it Column~8\rm : average logarithmic slope $\gamma$ of the density profile $\rho \propto r^{-\gamma}$ calculated using Eq. \ref{eq:gamma}.
\end{table*}

The distribution of mass in galaxies continues to be the subject of intense debate. The situation is clear at large radius: stellar kinematics, gas kinematics and gravitational lensing show that a dark matter of unknown nature dominates the gravitational potential of most galaxies (assuming Newtonian dynamics). In these regions the rotation curves are approximately flat and the total density profiles close to isothermal (\citealt{bosma1978,bosma1981,vanalbada1985,gavazzi2007,cappellari2015}, hereafter C15). Well inside galaxies' stellar body the situation is much more diverse but, in these regions, baryons and dynamics are closely connected \citep{sancisi2004,swaters2012,lelli2013}: in low-surface-brightness galaxies, rotation curves rise slowly and dark matter dominates the potential \citep{deblok2001}; in high-surface-brightness galaxies, rotation curves rise fast to approximately (or slightly above) their flat part, and baryons constitute most of the mass (\citealt{vanalbada1985}; \citealt{kent1987}; \citealt{sackett1997}; \citealt{palunas2000}; \citealt{cappellari2006,cappellari2013a}, hereafter C13; \citealt{noordermeer2007}).

This situation is reflected in the observed correlation between galaxies' circular velocity \vcirc\ measured at the largest possible radius and the stellar velocity dispersion $\sigma$ measured in the central regions of the stellar body (\citealt{whitmore1979,gerhard2001,ferrarese2002,baes2003,pizzella2005}, hereafter P05; \citealt{courteau2007}, hereafter C07; \citealt{ho2007}; \citetalias{cappellari2013a}). In particular, the slope of this correlation depends on galaxy surface brightness in a way that, to first order, can be linked to the shape of the rotation curve (\citetalias{courteau2007}; \citealt{kormendy2011}). In simple terms, $\sigma$ traces the inner gravitational potential and can be related to the circular velocity measured well within the stellar body. If the rotation curve has already reached its flat part in the regions where $\sigma$ is measured (as in high-surface-brightness disc galaxies) then $v_\mathrm{circ}\sim1.4\ \sigma$, close to the theoretical expectation for an isothermal density profile. However, if the rotation curve is still rising (as in low-surface-brightness galaxies) $\sigma$ is lower and the \vcirc-$\sigma$ relation is steeper, while if the rotation curve declines at large radius (as in early-type spirals) the relation should be shallower.

Rotation curves and the correlation between \vcirc\ and $\sigma$ are powerful tools to investigate the relative distribution of baryons and dark matter in galaxies of different type, provided that \vcirc\ is measured all the way to the dark-matter dominated outer regions. This is relatively straightforward for late-type galaxies, where the abundant and dynamically-cold \hi\ traces the potential well outside the stellar body. Indeed, for such objects, resolved \hi\ observations have been used in this field for decades \citep[e.g.,][]{bosma1978,begeman1991,verheijen2001b,martinsson2016}.

In contrast, in early-type galaxies (ellipticals and lenticulars; hereafter ETGs) \hi\ is detected less frequently and usually other methods must be used to measure \vcirc. Complex modelling of the stellar kinematics is generally required but this is typically limited to relatively small radius, comparable to the half-light radius \re\ (\citealt{kronawitter2000,gerhard2001,wegner2012}; \citetalias{cappellari2013a}). The resulting rotation curves and \vcirc-$\sigma$ relations apply therefore only to the baryon-dominated regions of these galaxies and do not cover the transition to the dark-matter dominated outskirts. As such, they provide limited information about the relation between baryons and dark matter in ETGs. Apart for a few individual objects \citep[e.g.][]{napolitano2009,napolitano2014,weijmans2009,murphy2011}, the only notable exception is the study of \citetalias{cappellari2015}, whose dynamical models of 14 ETGs reach a median radius of 4 \re.

Among the above studies, only \cite{ho2007} provides measurements of \vcirc\ for a large number of ETGs based on \hi\ data, potentially probing the dark-matter dominated regime. However, these values are derived from unresolved \hi\ spectra obtained with single dish telescopes, and the uncertainty on the dynamical state and geometry of the detected gas results in no significant correlation between \vhi\ and $\sigma$. More reliable, resolved \vhi\ measurements at large radius are available for a handful of ETGs included in the samples of \citetalias{pizzella2005} and \citetalias{courteau2007}. However, both samples are dominated by galaxies whose \vcirc\ is estimated from stellar kinematics and dynamical modelling at small radius. Neither \citetalias{pizzella2005} nor \citetalias{courteau2007} derive a \vcirc-$\sigma$ relation for ETGs based on resolved \hi\ data alone. We discuss these studies in more detail in Sec. \ref{sec:resmaster}.

Here we take a step forward by using new, interferometric \vhi\ estimates obtained at a median radius of 6 \re\ by \citet[][hereafter dH15]{denheijer2015} for a sample of 16 ETGs. Thanks to the size of this sample we are able to establish a tight linear \vcirc-$\sigma$ relation for ETGs using solely \vcirc\ values derived within the dark-matter dominated regime from resolved \hi\ data. We further combine our \vhi\ estimates with models by \citetalias{cappellari2013a} to study the typical shape of the rotation curve in these objects, and measure the average slope of their density profiles out to large radius.

\section{Sample and Data}
\label{sec:sample}

We analyse a sample of 16 nearby ETGs drawn from the \atl\ sample \citep{cappellari2011a}. All galaxies and quantities used for this work are listed in Table \ref{table:tab1}. Galaxies are selected for hosting a regular \hi\ disc or ring, which allows the determination of \vhi\ at large radius. The selection is based on a large set of interferometric \hi\ observations and data products presented by \cite{serra2012a,serra2014} for $\sim150$ galaxies. This dataset was largely assembled as part of the \atl\ project but includes earlier data taken by \cite{morganti2006}, \cite{weijmans2008}, \cite{jozsa2009} and \cite{oosterloo2010}. The \hi\ kinematics was modelled by \citetalias{denheijer2015} for all but two galaxies for which literature values were used: NGC~2685 \citep{jozsa2009} and NGC~2974 \citep{weijmans2008}. Here we make use of the \vhi\ values measured at the largest possible radius \rhi\ and listed by \citetalias{denheijer2015}. The \rhi\ values fall in the range 4 to 16 \re, and their median is $\sim6$ \re.

All galaxies in the sample were observed with optical integral-field spectroscopy as part of the SAURON project (4 galaxies; \citealt{dezeeuw2002}) or \atl\ (12 galaxies; \citealt{cappellari2011a}). Here we make use of the stellar velocity dispersion measurements \se\ within a 1 \re\ aperture given in \citetalias{cappellari2013a}. The same paper describes Jeans anisotropic modelling of the stellar kinematics of these galaxies (JAM; \citealt{cappellari2008}).  Here use the maximum circular velocity \vjam\ values predicted by the JAM models at the radius \rmax. For this work we use JAM models (A), which assume that mass follows light and are appropriate for the central regions of galaxies. The values of \rmax\ fall in the range 0.1 to 0.4 \re\ (median $\sim0.2$ \re).

This sample does not have any strong bias on the mass-size plane of ETGs, and the 16 galaxies cover a representative range of bulge-to-disc ratio (as traced by \se) in the stellar mass range $10^{10}$ - $10^{11}$ M$_\odot$ \citepalias{denheijer2015}. Bearing in mind the small sample size, the conclusions of this work should hold for the general early-type population within this mass range and with \se\ between 100 and 250 \kms.

\section{Results}
\label{sec:resmaster}

\subsection{Linear relation between \vhi\ and \se}
\label{sec:res1}

\begin{figure}
\centering
\includegraphics[width=8.5cm]{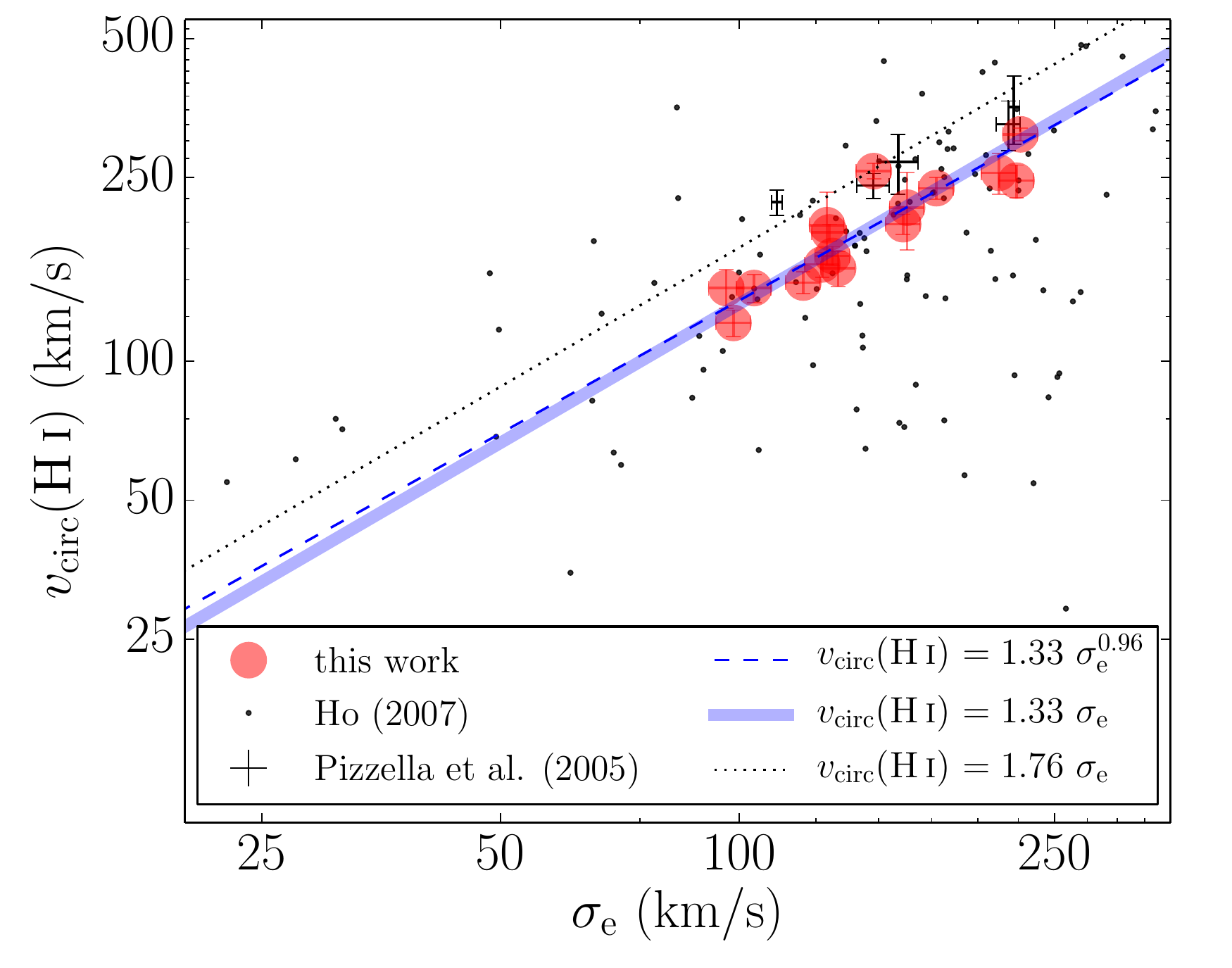}
\caption{Relation between \vhi\ and \se. Dashed and solid lines show the best-fitting power-law and linear relation, respectively, obtained for our sample of 16 ETGs (see Sec. \ref{sec:res1} and legend). The dotted line is a linear fit (with zero intercept) to the P05 data, noting that for consistency with our work we correct their $\sigma$ measurements to an aperture of radius 1 \re\ (see Sec. \ref{sec:comp} for details). We do not correct the $\sigma$ values of Ho (2007) because that sample shows no correlation between \vhi\ and $\sigma$ in the first place (Sec. \ref{sec:comp}), and the correction is irrelevant for the purpose of our comparison.}
\label{fig:mainresult}
\end{figure}

We plot \vhi\ against \se\ in Fig. \ref{fig:mainresult}. The figure shows a strong correlation between these two quantities. A power-law fit to our data performed as a linear fit in logarithmic space with the \texttt{LTS\_LINEFIT} software \citepalias{cappellari2013a} results in:

\begin{equation}
\begin{split}
\log{\frac{v_\mathrm{circ}(\mathrm{\mbox{\hi}})}{\mathrm{km \ s^{-1}}}} & = (2.299\pm0.013) + \\ 
& + (0.96\pm0.11) \times \log{\frac{\sigma_\mathrm{e}}{150 \ \mathrm{km \ s^{-1}}}},
\label{eq:powerlaw}
\end{split}
\end{equation}

\noindent where the value 150 \kms\ was adopted to minimise the covariance between the fit parameters (\citetalias{cappellari2013a}). The observed r.m.s. scatter around this relation is 0.049 dex (12 percent). The intrinsic scatter is $0.036\pm0.016$ dex ($9\pm4$ percent).

The value of the power-law exponent in Eq. \ref{eq:powerlaw} is consistent with unity. Indeed, Fig. \ref{fig:mainresult} shows that, within the \se\ range of our sample, the best-fitting power-law is fully compatible with the linear relation:

\begin{equation}
v_\mathrm{circ}(\mathrm{\mbox{\hi}}) = 1.33 \  \sigma_\mathrm{e},
\label{eq:linear}
\end{equation}

\noindent where the uncertainty on the slope is 3 percent. The scatter around this relation is 12 percent, identical to the scatter around the best-fitting power law. This is the first time that such a tight, linear relation is found using \vcirc\ measurements obtained at such a large radius ($\sim6$ \re) from the resolved \hi\ kinematics of a sizeable sample of ETGs. In Sec. \ref{sec:comp} we support this claim by performing a detailed comparison between this result and previous work.

The existence of a correlation between \vhi\ and \se\ for ETGs is expected given that  $\sigma$ correlates with $L_\mathrm{K}$ \citep{faber1976} and $L_\mathrm{K}$ correlates with \vhi\ (\citetalias{denheijer2015}). However, the actual linearity, slope and scatter of the relation, which we establish here, are not trivial. In particular, its linearity implies that the $L_\mathrm{K}$-$\sigma$ and $L_\mathrm{K}$-\vhi\ power-law relations have identical exponents, in agreement with general predictions of the Modified Newtonian Dynamics \citep{milgrom1984}.

Importantly, the \vhi-\se\ relation is relatively free of systematics as  both quantities involved are independent of distance and stellar mass-to-light ratio (unlike in other galaxy scaling relations), and have small errors. The main source of uncertainty on \vhi\ is the inclination of the gas disc, which results in a typical error below 10 percent (\citetalias{denheijer2015}; Table \ref{table:tab1}). The uncertainty on \se\ is $\sim5$ percent \citepalias{cappellari2013a}.

\begin{figure}
\centering
\includegraphics[width=8.5cm]{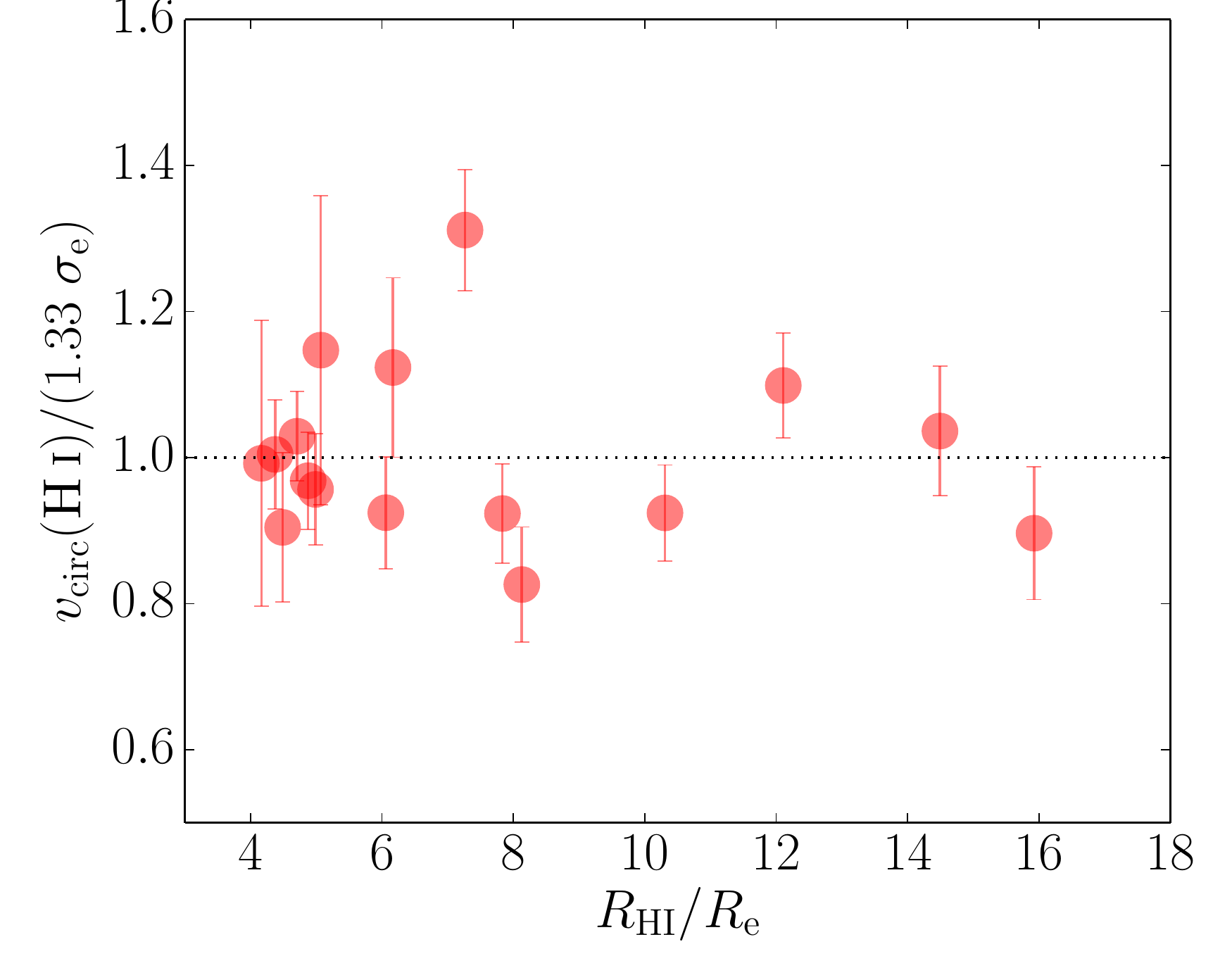}
\caption{Deviation from the \vhi-\se\ relation of Eq. \ref{eq:linear} as a function of \rhi/\re. No relation is found.}
\label{fig:deviations}
\end{figure}

The small intrinsic scatter of the \vhi-\se\ relation is particularly interesting. In Secs. \ref{sec:comp} and \ref{sec:res2} we discuss evidence that \vcirc\ declines with radius in ETGs, and this raises the question of whether some of this scatter is related to the relatively large range of \rhi\ values covered by our sample. We investigate this aspect in Fig. \ref{fig:deviations}. The figure shows no clear trend between the deviation from the relation and \rhi/\re: galaxies where \vhi\ is measured at larger (lower) radius are not systematically below (above) the relation. This suggests that other factors drive the observed scatter. An important contribution might come from the scatter in the shape of the ETG rotation curves. This can be linked to the scatter in the mass-concentration relation of dark matter halos predicted by dark-matter-only simulations \citep{dutton2014}. In the future it will be interesting to test whether the intrinsic scatter of the \vhi-\se\ relation is consistent with these simulations (see \citealt{dutton2012} and \citealt{lelli2016} for a similar test based on the intrinsic scatter of the baryonic Tully-Fisher relation). 

The relation between \vhi\ and \se\ is remarkable as the two quantities are measured in completely independent ways and, unlike in previous work, trace the gravitational potential in two extremely different regimes of the distribution of matter: \se\ is measured in a region where the luminous matter constitutes on average 85 percent of the mass \citepalias{cappellari2013a}; while \vhi\ is measured in a region where dark matter accounts for basically all of the mass (the total dynamical mass within \rhi\ is on average $\gtrsim 3$ times the total stellar mass estimated by \citetalias{cappellari2013a}). This tight correlation suggests that galaxies in our sample all have roughly the same overall distribution of mass -- another manifestation of the poorly understood coupling between the distribution of luminous matter and total mass in galaxies \citep{vanalbada1986,sancisi2004,swaters2012,lelli2013}. We will return to this point in Sec. \ref{sec:res3}.

\subsection{Comparison with previous \vcirc-$\sigma$ relations}
\label{sec:comp}

\subsubsection{\vcirc\ from \hi\ kinematics}
\label{sec:hicomp}

The only previous study of the \vcirc-$\sigma$ relation of ETGs entirely based on \hi\ data is that of \cite{ho2007}, whose sample includes $\sim100$ ellipticals and lenticulars. However, that work makes use of single-dish unresolved \hi\ data, which give no information about the dynamical state and geometry of the gas. In order to derive \vcirc\ from unresolved spectra one must assume that \it i)~\rm the \hi\ is on a rotating disc and \it ii)~\rm  the disc has the same inclination of the stellar body.  These two assumptions are in error for more than half of all \hi-rich ETGs \citep{serra2012a,serra2014}, resulting in inaccurate single-dish \vhi\ values for this type of galaxies. It is likely for this reason that \vhi\ and $\sigma$ do not correlate significantly for the ETG sample of \cite{ho2007}, as shown in Fig. \ref{fig:mainresult}. We have independently obtained a Spearman correlation coefficient of 0.41 (with p-value 0.17) for that sample, confirming the visual impression given by the figure.

\cite{ho2007} finds that \vhi\ and $\sigma$ do correlate if they limit their analysis to a subsample of ETGs selected to follow the \vhi-$L_\mathrm{K}$ (Tully-Fisher) relation of spirals. However, we argue that for such a subsample the \vhi-\se\ relation is a selection effect. A Tully-Fisher selection will always result in a \vhi-$\sigma$ relation -- even if none existed and regardless of the quality of the \vhi\ estimates -- because of the tight  $\sigma$-$L_\mathrm{K}$ (Faber-Jackson) relation of ETGs. Furthermore, the shape and scatter of the resulting \vhi-$\sigma$ relation will be entirely determined by the details of the Tully-Fisher selection rather than by the properties of the galaxies being studied. Our galaxies, too, follow the Tully-Fisher relation (\citetalias{denheijer2015}), but they are not selected to do so. Therefore, the existence, shape and scatter of our \vhi-$\sigma$ relation fully reflect the properties of the ETGs in our sample rather than the way they are selected.

The only previous studies of the \vcirc-$\sigma$ relation to overcome the limitations of single-dish data and include resolved \vhi\ measurements for ETGs are those by \citetalias{pizzella2005} and, using the same \hi\ data, \citetalias{courteau2007}. The main difference from our analysis is that the number of \hi\ measurements used is very small -- just 5. Apart from these 5 objects, those ETG samples are dominated by \vcirc\ values obtained at a radius comparable with \re\ from stellar kinematics and dynamical modelling (discussed in Sec. \ref{sec:stellarcomp}), and these authors do not derive an ETG \vcirc-$\sigma$ relation based on \hi\ data alone. Nevertheless, in this Section we discuss whether the tight linear relation \vhi~=~1.33~\se\ obtained in Sec. \ref{sec:res1} from our data could have been derived from the 5 ETGs with  \hi\ data included in the \citetalias{pizzella2005} and \citetalias{courteau2007} samples.

Fig. \ref{fig:mainresult} shows those 5 ETGs on the \vhi-\se\ plane. An important detail is that \citetalias{pizzella2005} and \citetalias{courteau2007} use $\sigma$ measurements obtained within (or corrected to) an $R_\mathrm{e}/8$ aperture. For consistency with our work we correct those values to a 1 \re\ aperture multiplying them by a factor 0.872 derived from Eq. 1 of \cite{cappellari2006}. The accuracy of (and need for) this correction can be easily verified considering the two ETGs in common with our sample, NGC~2974 and NGC~4278.

As a whole, the 5 ETGs of \citetalias{pizzella2005} and \citetalias{courteau2007} appear systematically offset towards high \vhi\ relative to our sample in Fig. \ref{fig:mainresult}. Furthermore, they do not follow a linear \vhi-\se\ relation. The best-fitting power law obtained with the same \texttt{LTS\_LINEFIT} software used in Sec. \ref{sec:res1} has an exponent of $0.59\pm0.17$. This is significantly non-linear, unlike our Eq. \ref{eq:powerlaw}. Assuming linearity, the slope of the relation defined by those 5 ETGs and obtained with an orthogonal distance regression fit of a linear relation through the origin is 1.76 with a 6 percent uncertainty (Fig. \ref{fig:mainresult}). This is $\sim30$ percent larger than the 1.33 slope of our relation.

We note that, taken individually, none of the 5 ETGs studied by \citetalias{pizzella2005} and \citetalias{courteau2007} is dramatically offset from our sample in Fig. \ref{fig:mainresult}. The average $\sim30$ percent offset discussed above is most likely caused by the small number of objects combined with the large uncertainty on \vhi. More in detail, \citetalias{pizzella2005} adopt \vhi\ values obtained from data with a velocity resolution of $\sim40$ \kms\ for the two galaxies in common with our sample: NGC~2974 \citep{kim1988} and NGC~4278 (\citealt{lees1994}). Despite being consistent with our measurements within the large errors, the resulting \vhi\ values of $355\pm60$ and $326\pm40$ \kms\ obtained at $\sim4$ and $\sim10$\re, respectively, are 15 and 27 percent larger than those in Table \ref{table:tab1}. We also note that \cite{lees1994} estimate \vhi\ in NGC~4278 through complex modelling of the \hi\ data cube assuming a triaxial gravitational potential at constant inclination, which may introduce systematic differences compared to the method used by \citetalias{denheijer2015}. For NGC~2865 \citetalias{pizzella2005} adopt the \vhi\ estimate at $\sim6$\re\ from \cite{schiminovich1995}, who remark that their assumption of coplanar circular orbits for the \hi\ gas is not necessarily supported by their data. For NGC~5266 \vhi\ comes from a study by \cite{morganti1997}, where again circular orbits are just assumed. These authors estimate \vhi\ at $\sim4$\re\ while at larger radius the gas disc appears unsettled. Finally, for IC~2006 -- the main outlier in Fig. \ref{fig:mainresult} -- \vhi\ is taken from \cite{franx1994}. Their analysis shows that the rotation curve of this galaxy is relatively flat. As we discuss below, this is not typical for ETGs.

The above considerations highlight that it would have been difficult for \citetalias{pizzella2005} to obtain a reliable slope of the \vhi-\se\ relation based on such few objects and sometimes limited data quality. The inclusion of these 5 galaxies in their sample was important at the time, and that paper clearly shows that \vhi\ grows with \se\ in ETGs. However, that sample was just too small to establish the linearity, tightness and actual slope of the \vhi-\se\ relation presented in Sec. \ref{sec:res1}.

\subsubsection{\vcirc\ from stellar kinematics at small radius}
\label{sec:stellarcomp}

Having established a precise \vhi-\se\ relation for ETGs based on our data, we can investigate whether such relation is identical to those obtained using kinematical data at smaller radius. This was one of the conclusions of \citetalias{pizzella2005}: their 5 ETGs with a \vhi\ measurement follow the same \vcirc-$\sigma$ relation obtained for a larger sample consisting of 40 high-surface-brightness spirals (with \vcirc\ from ionised gas or \hi\ data; see references in \citetalias{pizzella2005}) and another 19 ETGs (with \vcirc\ measured within $\sim2$~\re\ from the dynamical models of \citealt{kronawitter2000}). Below we show that our data do not support this conclusion, and Eq. \ref{eq:linear} is markedly shallower than relations obtained at smaller radius.

First, assuming that the relation between \vcirc\ and $\sigma$ is linear, \citetalias{pizzella2005} find \vcirc $=(1.32\pm0.09)\ \sigma+46\pm14$. This relation is significantly different from our Eq. \ref{eq:linear} for two reasons. First, it has a non-zero intercept which cannot be ignored. At the median velocity dispersion of  our sample ($\sim130$ \kms), the intercept amounts to $\sim20$ percent of the \vcirc\ value predicted by the relation. Second, the $\sigma$ values used by \citetalias{pizzella2005} must be corrected to the 1 \re\ aperture used in our work (Sec. \ref{sec:hicomp}). When this is done, we find that the best-fitting slope of a linear relation through the origin for the full \citetalias{pizzella2005} sample is 1.77, $\sim30$ percent larger than our value of 1.33.

A similar conclusion can be drawn for other samples dominated by \vcirc\ measurements at small radius. Focusing on ETGs, \citetalias{courteau2007} adds to the \citetalias{pizzella2005} sample $\sim50$ lenticulars with \vcirc\ measured by \citet[][and references therein]{bedregal2006} based on stellar kinematics -- but no new \hi\ data. Once their $\sigma$ values are corrected to an aperture of 1 \re\ as above, these ETGs scatter about a \vcirc-\se\ relation of slope $\sim1.7$ and are clearly inconsistent with our Eq. \ref{eq:linear}.

Likewise, our slope of 1.33 is significantly lower than the 1.52 value found by \cite{gerhard2001}. The comparison with the latter study is particularly useful to understand the cause of this difference. First, as above, we correct for the larger aperture within which we measure $\sigma$ (1 \re\ compared to $\sim0.1$ \re). We use again Eq. 1 of \cite{cappellari2006} and find $\sigma_\mathrm{e}=0.859\ \sigma_{0.1R_\mathrm{e}}$. The other, crucial difference from this and all other previous studies of ETGs is that our \vcirc\ values are obtained at a much larger radius ($\sim6$ \re\ compared to $\sim0.3$ \re\ in the case of \citealt{gerhard2001}). This difference is relevant in light of the recent findings of \citetalias{cappellari2015}. They show that the density profiles of ETGs are somewhat steeper than isothermal, and that \vcirc\ decreases slowly with radius. We discuss this aspect in more detail in Secs. \ref{sec:res2} and \ref{sec:res3}, but for the moment it is sufficient to consider that on average $v_\mathrm{circ} \propto r^{-0.095}$. Therefore, our \vcirc\ measurements and those of \cite{gerhard2001} are related by $v_\mathrm{circ}(6R_\mathrm{e})=0.752\ v_\mathrm{circ}(0.3R_\mathrm{e})$. Taken together, the two corrections transform our Eq. \ref{eq:linear} from $v_\mathrm{circ}(6R_\mathrm{e})=1.33\ \sigma_\mathrm{e}$ into $v_\mathrm{circ}(0.3R_\mathrm{e})=1.52\ \sigma_{0.1R_\mathrm{e}}$, which is exactly the result of \cite{gerhard2001}.

Finally, \citetalias{cappellari2013a} derives the \vcirc-\se\ relation from dynamical models of all 260 ETGs in the volume-limited, complete \atl\ sample. They find a slope of 1.76 when using the peak \vcirc\ predicted by the models (typically at 0.2 \re). As for the study of \cite{gerhard2001}, the larger slope compared to Eq. \ref{eq:linear} can be explained by the decline of \vcirc\ with radius (\citetalias{cappellari2015} and Sec. \ref{sec:res2}).

In summary, once aperture effects on the measurement of $\sigma$ are taken into account, our \vhi-\se\ relation is shallower than all published relations based on \vcirc\ measurements at smaller radius. This suggests that \vcirc\ is not constant with radius and indicates the validity of the conclusions of \citetalias{cappellari2015} on the slow decline of \vcirc. It highlights that ETG samples selected to only include objects with a flat rotation curve (as in \citetalias{pizzella2005}) might be biased towards galaxies with a specific distribution of total mass, such that \vcirc\ remains high even when it is measured at relatively large radius. We explore the variation of \vcirc\ with radius in ETGs in the next sections.

\subsection{Variation of \vcirc\ with radius}
\label{sec:res2}

\begin{figure}
\centering
\includegraphics[width=8.5cm]{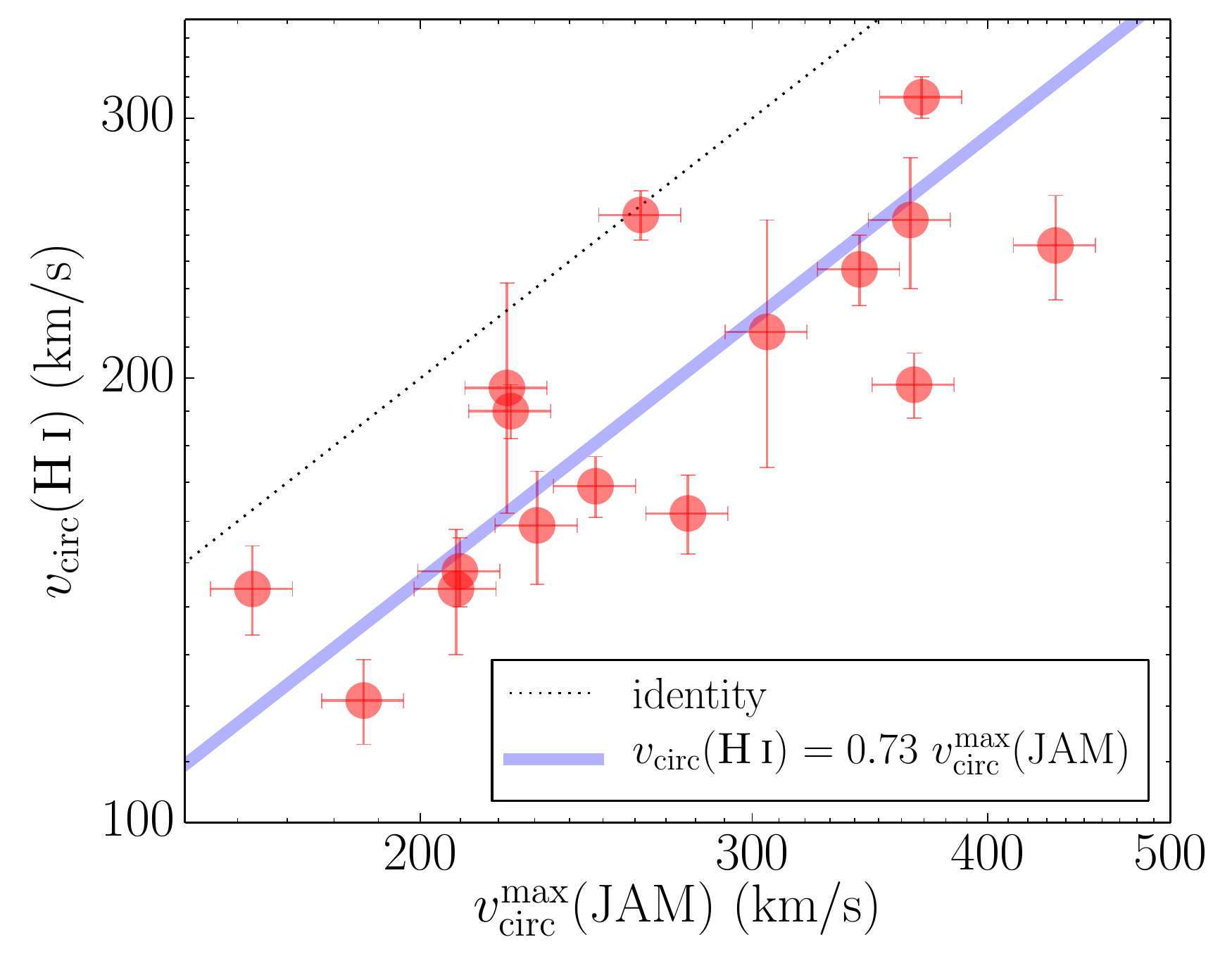}
\caption{Relation between \vhi\ and \vjam. The solid line shows the best-fitting linear relation (see Sec. \ref{sec:res2} and legend).}
\label{fig:mainresultJAM}
\end{figure}

We compare \vhi\ with the circular velocity at small radius \vjam\ derived by \citetalias{cappellari2013a} using dynamical models (see Sec. \ref{sec:sample} and Table \ref{table:tab1}). Those authors find a tight correlation between \vjam\ and \se. Therefore, in light of our Fig. \ref{fig:mainresult}, we expect a correlation between \vhi\ and \vjam. This is shown in Fig. \ref{fig:mainresultJAM}. As in Sec. \ref{sec:res1}, we fit a power law to the data points and find the best fit to be relatively close to a linear relation given the error bars (the exponent is $0.79 \pm 0.15$). Indeed, the 17 percent (0.068 dex) observed scatter around the best-fitting power law is only slightly lower than the 18 percent scatter around the linear relation:

\begin{equation}
v_\mathrm{circ}(\mathrm{\mbox{\hi}}) = 0.73 \  v_\mathrm{circ}^\mathrm{max}(\mathrm{JAM}).
\end{equation}

The scatter of the \vhi-\vjam\ relation is considerably larger than that of the \vhi-\se\ relation. In particular, based on the 7 percent observed scatter of the \vjam-\se\ relation (\citetalias{cappellari2013a}), a conservative estimate of the uncertainty on \vjam\ is $\sim5$ percent (as for \se). The resulting intrinsic scatter of the \vhi-\vjam\ relation is $16\pm5$ percent (it was $9\pm4$ percent for the \vhi-\se\ relation). This could indicate that, in the present sample, the distribution of mass is relatively homogeneous between \re\ and \rhi, but less so between \rmax\ and \re. A visual inspection of the rotation curves in the \cite{noordermeer2007} sample indicates that this is the case in early-type spirals.

Our measurements show that \vcirc\ drops on average \mbox{$\sim25$} percent from \rmax\ to \rhi\ (the actual drop varying between 0 and 50 percent depending on the galaxy). Obvious questions are where this drop occurs within galaxies and how gradual it is with radius. Some qualitative indications come from molecular gas \vcirc\ estimates obtained by \cite{davis2011a} at a radius intermediate between \rmax\ and \rhi\ (for CO-detected galaxies in the \atl\ sample the molecular gas reaches a typical radius between 0.5 and 1 \re; see \citealt{davis2013}). First, \cite{davis2011a} show that \vcirc(CO) is on average $\sim10$ percent lower than \vjam\ (with considerable scatter; see their fig. 4). Second, \citetalias{denheijer2015} report an offset between the CO and \hi\ $K$-band Tully-Fisher relations of ETGs corresponding to a \vcirc\ decrease by another $\sim10$ percent. Combining these two results we conclude that the average $\sim25$ percent \vcirc\ drop between \rmax\ and \rhi\ is about equally distributed inside and outside \re.

A more quantitative result would require a detailed study of the resolved rotation curve out to \rhi\ for galaxies in this sample. However, at the typical resolution of our \hi\ data ($\sim 40$ arcsec) this is possible for just a few objects. Two such galaxies are NGC~2685 \citep{jozsa2009} and NGC~2974 \citep{weijmans2008}. Their rotation curves show indeed a clear decline out to $\sim10$ and $\sim2$ \re, respectively, and appear to flatten further out. This would be consistent with the situation in early-type spirals, where \vcirc\ peaks at small radius and then drops by 10-20 percent to a flat level \citep{noordermeer2007}. A similar study of the resolved \hi\ rotation curve can be performed for a few more objects in our sample and will be the subject of future work. For the time being, it remains unclear whether and at what characteristic radius the \hi\ rotation curves of ETGs become flat.

Whatever results will be obtained from studying the resolved \hi\ kinematics of more ETGs, we do know from \citetalias{cappellari2015} that the stellar rotation curves of most such systems decline steadily out to at least $\sim4$ \re. These authors find that the observed decline of \vcirc\ is well described by power-law density profiles $\rho \propto r^{-\gamma}$, and that the profile slopes $\gamma$ cover a surprisingly narrow range centred around a mean value $\langle\gamma\rangle= 2.19\pm0.03$ and with an observed r.m.s. scatter $\sigma_\gamma$ of just 0.11. Here we can test this result using a different sample (only two galaxies are in common between our sample and that of \citetalias{cappellari2015}) and completely different observations that reach further out into the dark-matter halo.

\begin{figure}
\centering
\includegraphics[width=8.5cm]{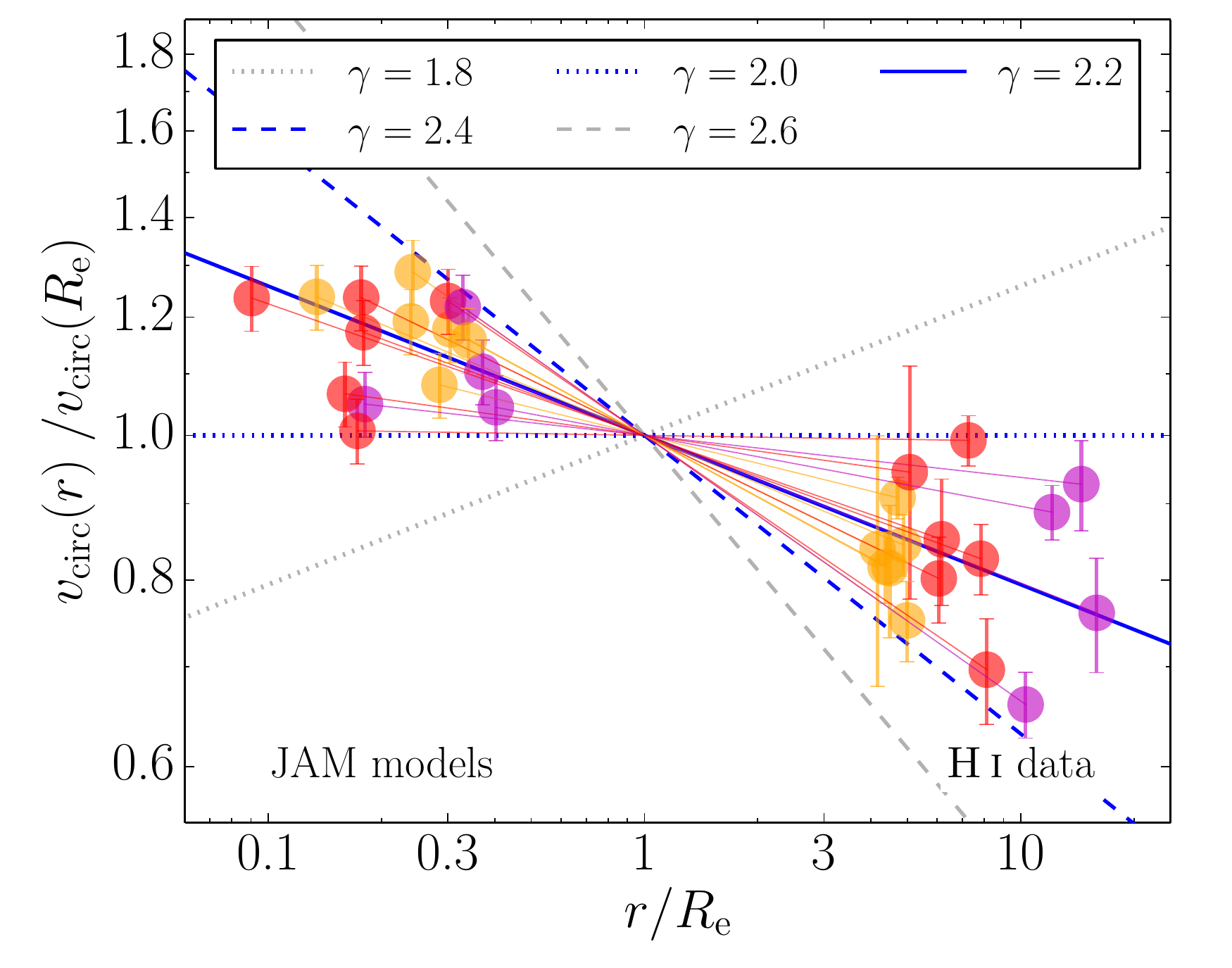}
\caption{Variation of \vcirc\  with radius based on estimates at \rmax<\re\ from JAM models and at \rhi>\re\ from \hi\ data. Points and connecting solid lines are colour coded in three groups going from orange through red to magenta according to increasing \rhi/\re\ ratio. For each galaxy, the values of \vcirc\ at \rmax\ and \rhi\ are normalised to \vcirc(\re), but note that we do not make use of actual \vcirc(\re) measurements. Instead, for normalisation purpose, we assume that \vcirc(\re) lies on the rotation curve passing through our measurements at \rmax\ and \rhi\ with logarithmic slope defined by Eq. \ref{eq:gamma}. Blue and grey lines correspond to models with a power-law density profile $\rho \propto r^{-\gamma}$ for the values of $\gamma$ listed in the legend.}
\label{fig:gamavel}
\end{figure}

\subsection{Average slope of the total density profiles}
\label{sec:res3}

The analysis of \citetalias{cappellari2015} shows that the mass distribution of ETGs is well represented by power-law density profiles $\rho \propto r^{-\gamma}$ out to large radius. We therefore assume power law profiles  for our galaxies. It follows that $v_\mathrm{circ} \propto r^{1-\gamma/2}$ \citep[][Eq. 2.61]{binney2008}. We can then use our measurements of \vjam\ and \vhi\ at \rmax\ and \rhi, respectively, to measure the average logarithmic slope of the density profile for each galaxy:

\begin{equation}
\gamma = 2 - 2 \ \frac{\log{v_\mathrm{circ}(\mathrm{\mbox{\hi}})}-\log{v_\mathrm{circ}^\mathrm{max}(\mathrm{\mbox{JAM}})}}{\log{R_\mathrm{HI}}-\log{R_\mathrm{max}}},
\label{eq:gamma}
\end{equation}

\noindent and study the distribution of $\gamma$ values across the sample.

Fig. \ref{fig:gamavel} shows our measurements as well as model rotation curves for a range of $\gamma$ values. We find a narrow range for $\gamma$ , with all galaxies confined in a region of the plot corresponding to $2<\gamma<2.4$. On average, galaxies in this sample appear to have profiles somewhat steeper than isothermal. This is an important confirmation of the result presented by \citetalias{cappellari2015}, in particular considering the different sample and type of data used here.

More quantitatively, the weighted mean and the r.m.s. scatter of the 16 $\gamma$ values listed in Table \ref{table:tab1} are:

\begin{equation}
\langle\gamma\rangle=2.18\pm0.03 \ \ \mathrm{and} \ \ \sigma_\gamma=0.11 \ .
\end{equation}

\noindent Although here we cannot check whether the assumption of power-law profiles is correct and can only measure the average logarithmic slope, our result is in remarkable quantitative agreement with the one of \citetalias{cappellari2015}, where the adequacy of power-law profiles is verified. Our mean slope is very close to their $\langle\gamma\rangle=2.19\pm0.03$, and the observed scatter is identical. Our result is also in relatively good agreement with that obtained at small radius from gravitational lensing, which suggests profile shapes only marginally closer to isothermal \citep{auger2010,barnabe2011}. 

As discussed in Sec. \ref{sec:res2}, our data do not rule out that at some radius beyond \re\ the rotation curves of ETGs become flat. If that is the case, we would expect our measurements of $\gamma$ to approach the isothermal value $\gamma=2$ as \rhi\ increases. Fig. \ref{fig:gamavel} does not show any clear indications that this is the case. Galaxies with a \vhi\ measurement at $\sim10$ \re\ or above can have a value of $\gamma$ significantly above 2, while galaxies with a \vhi\ measurement obtained below $\sim5$ \re\ can have $\gamma$ close to the isothermal value. As pointed out in the case of the \vhi-\se\ relation (Sec. \ref{sec:res1} and Fig. \ref{fig:deviations}), the scatter in the shape of the ETG rotation curves might be more important than the radius at which we measure \vhi. We stress again that only the resolved study of more \hi\ rotation curves can clarify whether and at what radius \vcirc\ becomes flat in ETGs.

\begin{figure}
\centering
\includegraphics[width=8.5cm]{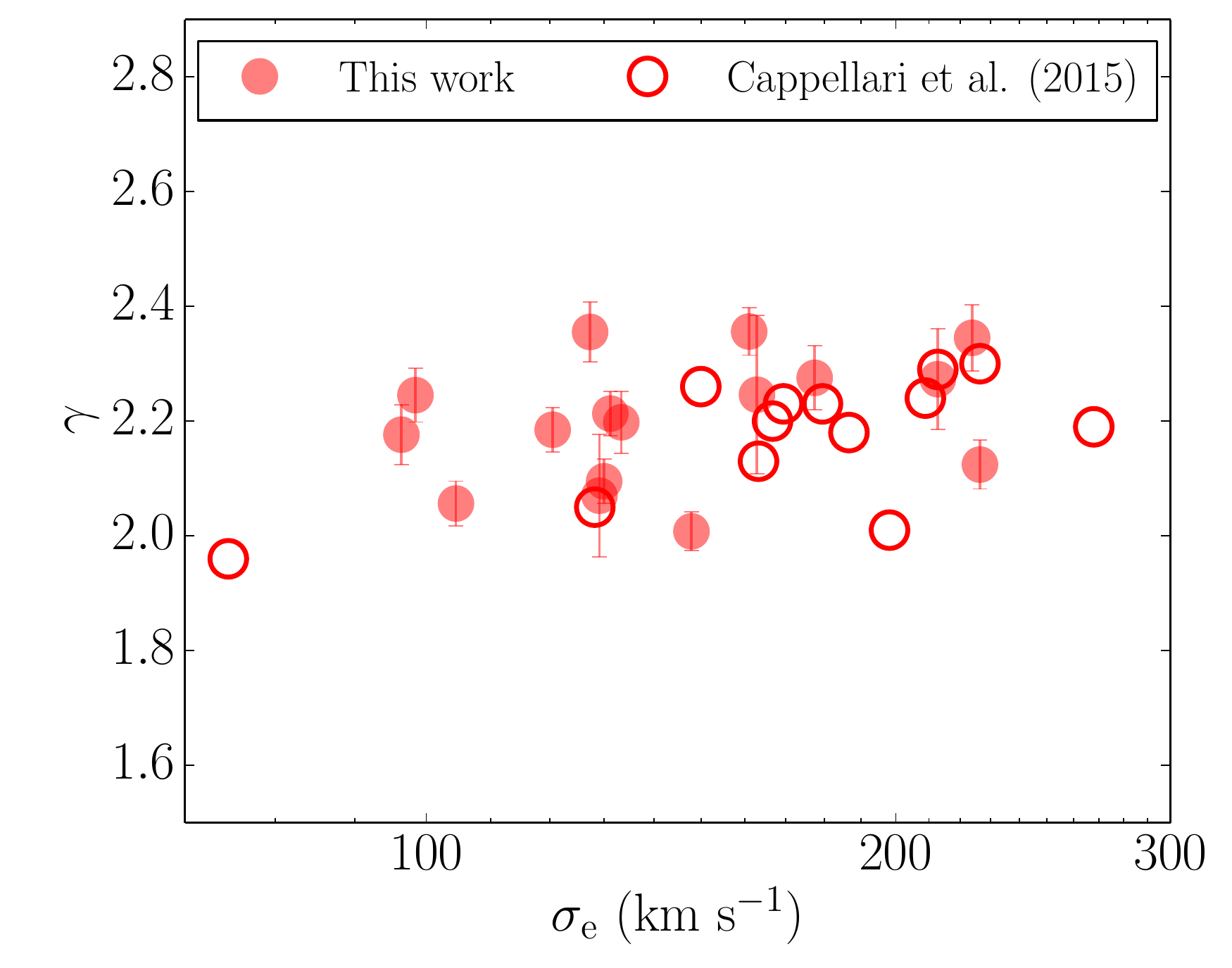}
\caption{Average logarithmic profile slope $\gamma$ vs. stellar velocity dispersion \se\ for our sample and the sample of \citetalias{cappellari2015}. No correlation is observed.}
\label{fig:gasigma}
\end{figure}

The above discussion indicates that, whether or not the density profiles of all ETGs are well approximated by single power-laws out to the radius probed by our \vhi\ measurements, \it (i) \rm their average slopes cluster around a value of 2.2 and \it (ii) \rm the small scatter of 0.11 is indicative of relatively small differences between galaxies, as also suggested by the scatter of the \vhi-\se\ relation. The origin of these differences is, however, unclear. Since $\gamma$ measures an average property of the distribution of mass one may think that its value depends on the presence of a massive bulge and, therefore, on the value of \se. However, Fig. \ref{fig:gasigma} shows that $\gamma$ does not vary systematically with \se\ for our sample combined with the one of \citetalias{cappellari2015} (the correlation coefficient is below 0.4). An identical conclusion is reached from dynamical modelling of ETGs within 1 \re\ considering only galaxies within the \se\ and stellar mass range covered by our sample (fig. 22 of \citealt{cappellari2016}). The same modelling shows that $\gamma$ does decrease at lower \se\ and larger stellar masses than covered by our sample, but it is unknown whether a similar trend persists at the large, dark-matter dominated radii reached by our \hi\ data.

We have investigated possible correlations between $\gamma$ and additional parameters that trace the relative importance of the bulge: the bulge-to-total ratio, the effective surface brightness (both total and of the bulge) and the Sersic index from \cite{krajnovic2013a}; the light concentration from \citetalias{cappellari2013a}; and the ratio $\sigma_{R_\mathrm{e}/8}/\sigma_\mathrm{e}$ from \cite{cappellari2013b}. In all cases we find correlation coefficients below 0.4 (in absolute value). Therefore, we conclude that the current sample shows no indication of a systematic variation of $\gamma$ with the optical structure of ETGs. Larger samples will be needed to establish the nature of the small scatter of the density profile shapes.

\section{Conclusions}

We establish a tight linear relation between the circular velocity measured in the dark-matter dominated regions ($\sim6$ \re) and the velocity dispersion measured inside 1 \re\ for a sizeable sample of 16 ETGs. The \vcirc\ values are obtained from resolved \hi\ observations and, therefore, do not suffer from the limitations of single-dish data previously used in the literature. We find that \mbox{\vhi\ $=1.33$ \se} with an observed scatter of 12 percent. The tightness of the correlation suggests a strong coupling between luminous and dark matter from the inner regions where we measure \se\ to the outer regions where we measure \vhi. This coupling has been long known for spirals and dwarf irregulars \citep{vanalbada1986,sancisi2004,swaters2012,lelli2013} but had never been established for ETGs.

Previous samples of ETGs with resolved \vhi\ measurements were too small to establish the tightness, linearity and actual slope of the \vhi-\se\ relation, and to compare it with relations obtained at smaller radius. Overall, we find that our relation is shallower than those based on \vcirc\ measurements obtained from stellar kinematics and dynamical modelling at a radius comparable with \re. This indicates a decline in \vcirc\ from \re\ to the outer regions where we measure \vhi.

Comparing our \hi\ measurements of \vcirc\ at $\sim6$ \re\ with those derived from dynamical modelling of the stellar kinematics at much smaller radius ($\sim0.2$ \re) we find that the rotation curves of ETGs drop by 0 to 50 percent towards large radius for galaxies in the sample (25 percent on average). This drop is similar to that observed in early-type spirals \citep{noordermeer2007}. It is in excellent agreement with a recent, independent determination of the slope of the density profile of ETGs based on stellar kinematics and dynamical modelling of a different sample \citepalias{cappellari2015}. It appears that the average density profile of ETGs in the stellar mass and \se\ ranges probed by our combined samples is slightly steeper than isothermal, with a logarithmic slope of $2.19\pm0.02$ and a scatter of just 0.11.




\bibliographystyle{mnras}




\bsp	
\label{lastpage}
\end{document}